\documentstyle[12pt]{article}
\input{psfig}
\def\ang{\,{\rm\AA}}

\def\Mpc{\,{\rm Mpc}}

\def\cmm2{{\,\rm cm^{-2}}}
\def\cm2{{\,{\rm cm}^2}}
\def\cmm3{{\,{\rm cm}^{-3}}}
\def\gcmm3{{\,{\rm g\,cm^{-3}}}}
\def\kms{\,{\rm km\,s^{-1}}}

\def\fun#1#2{\lower3.6pt\vbox{\baselineskip0pt\lineskip.9pt
  \ialign{$\mathsurround=0pt#1\hfil##\hfil$\crcr#2\crcr\sim\crcr}}}

\begin{document}
\baselineskip=24pt
\pagestyle{empty}
\begin{center}
\bigskip

\rightline{astro-ph/9703160}
\rightline{published in {\it Nature} {\bf 381}, 193 (1996)}

\vspace{.2in}
{\Large \bf DEUTERONOMY AND NUMBERS}
\bigskip

\vspace{.1in}
David N. Schramm and Michael S. Turner\\

\end{center}

\vspace{.2in}
\setcounter{page}{1}

Four light isotopes -- D, $^3$He, $^4$He and $^7$Li --
were produced by nuclear reactions a few seconds
after the big bang.  New measurements of $^3$He in the ISM by Gloeckler
and Geiss (described on page xx of this issue) and of deuterium
in high redshift hydrogen clouds by Tytler and his collaborators
(described on page xx of this issue) provide further confirmation of
big-bang nucleosynthesis and new insight
about the density of ordinary matter (baryons).

Helium-4 is produced in the greatest abundance, by mass about
24\%.  The large primeval abundance of $^4$He, between 22\% and
25\%, provided the first confirmation of big-bang nucleosynthesis.
The other light elements are produced in much smaller amounts,
D and $^3$He around $10^{-5}-10^{-4}$ relative to hydrogen
and $^7$Li around $10^{-10}$ relative to hydrogen, but
play important roles in testing the theory.

The big-bang production of $^4$He is relatively insensitive to the density
of matter.   The yields of the other three light nuclei are
much more sensitive to the density and have the potential to
not only test the theory, but provide information about the
mean density of ordinary matter in the Universe (see Figure).

\begin{figure}[t]
\centerline{\psfig{figure=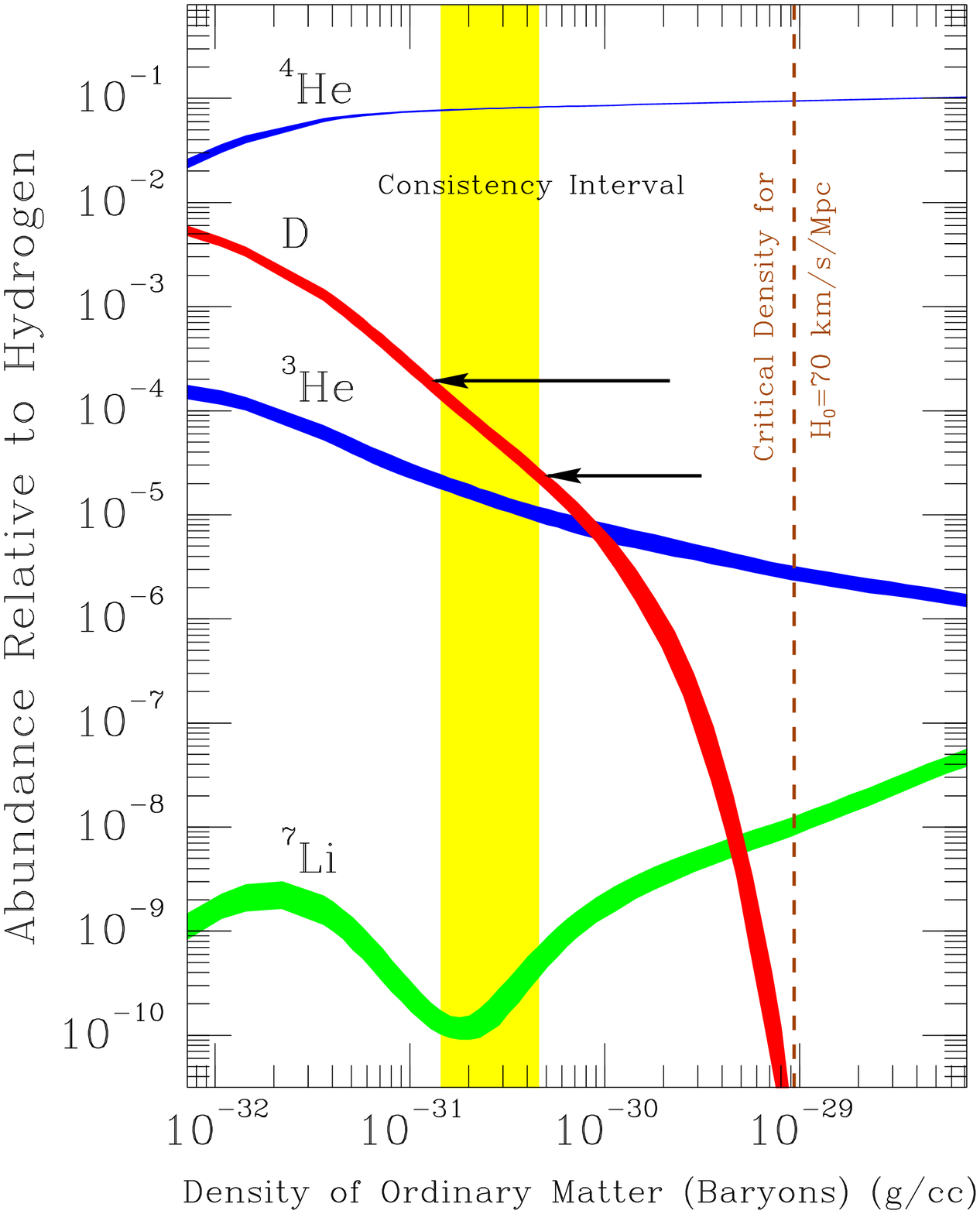,width=5in}}
\caption{Summary of big-bang production
of the light elements.  The widths of the curves
indicate the $2\sigma$ theoretical uncertainties, and the vertical
band is the consistency interval where the abundances of
all four light elements agree with their measured primeval
abundances.  The critical density, which depends upon the
square of the Hubble constant, is indicated for $H_0=70\kms\Mpc^{-1}$.
The arrows indicate the deuterium abundances measured by
Tytler et al. and Songaila et al.$^4$  (Figure courtesy of C.~Copi.)}
\end{figure}

The great success of primordial nucleosynthesis is that it
can account for the primeval abundances
of all four light elements (relative to hydrogen)
provided the baryon density
is between $1.5\times 10^{-31}\gcmm3$ and $4.5\times 10^{-31}\gcmm3$.
The consistency of the light-element abundances both
tests the big-bang cosmological model to within a fraction of a
second of the bang and provides the best ``measurement'' of
the density of ordinary matter$^1$.

Because big-bang deuterium production decreases rapidly with baryon density
and its post big-bang history
(or chemical evolution) is simple -- stars only destroy D -- it is
the best baryometer$^2$.
In the 1970's the first measurements of the D abundance in
the solar system and in the local interstellar medium
indicated that matter in the form of
baryons could account for at most 15\% of the critical density; this
conclusion still holds today.  Because of possible depletion
in stars, the present day deuterium abundance provides a lower limit
to the primeval value and thereby an upper limit to the baryon density.
(Terrestrial measurements of deuterium are of little cosmological use
because of severe isotopic fractionation effects.)

In 1976 Adams$^3$ proposed a way of measuring the
deuterium abundance in
very primitive samples of the cosmos where stellar depletion
would not be a concern.  From such a measurement the baryon
density could be inferred to an accuracy of 15\% or so.
He suggested that the deuterium Ly-$\alpha$
line, which in the rest frame is only $0.33\ang$
blueward of the usual $1216\ang$ Ly-$\alpha$ line,
might be detected in the wings of Ly-$\alpha$ absorption of
high-redshift hydrogen clouds backlit by distant QSOs.

For very distant clouds the UV lines of the Lyman series
of hydrogen lines
are shifted into the visible spectrum where they can
be detected with ground-based instruments.  But still,
this is no mean feat.  Isolated hydrogen
clouds of high column density and small velocity dispersion
have to be found if the much weaker deuterium feature is to be detected.
The very high spectral resolution and signal-to-noise ratio needed
demands the best instruments and largest telescopes.

With the advent of the 10\,meter Keck Telescope with its high-resolution
echelle spectrograph (HIRES) the dream is becoming reality.  In the past eighteen months
three detections and four tentative detections have been reported.
The hydrogen clouds studied are old -- redshifts between 2.6 and
4.7 -- and pristine -- heavy-element abundances ranging between $10^{-2}$
and less than $3\times 10^{-4}$ of that in the solar system.
(Heavy elements are made by stellar processes, not in the big bang.)
The values measured for the deuterium abundance fall into the range
anticipated and provide a dramatic confirmation of the big-bang prediction.

The two clouds studied by Tytler and his collaborators
have well determined
D abundances, (D/H) $=(2.3\pm 0.3)\times 10^{-5}$ and $(2.5\pm 0.3)\times
10^{-5}$, and indicate a baryon density of $(4.4\pm 0.6)\times
10^{-31}\gcmm3$.  The cloud
studied by Songaila and her collaborators$^4$
and Carswell and his collaborators$^5$ has a much higher abundance,
(D/H) $=(2 \pm 0.4)\times 10^{-4}$, and indicates
a baryon density of $(1.3\pm 0.3)\times 10^{-31}\gcmm3$.
The four tentative detections fall in between.

Any deuterium detection is most conservatively interpreted as an upper limit to
the primeval abundance because of the possibility of a low-column-density
hydrogen cloud fortuitously located to
mimic the deuterium feature (an interloper).  Thus, the feature
indicating a high value of deuterium might be a hydrogen
interloper, with the low value of deuterium reflecting the primeval
abundance.  If this is the case
the baryon density is at the high end of the range anticipated,
around 5\% of the critical density for a Hubble constant of $70\kms\Mpc^{-1}$.

However, Rugers and Hogan$^6$ estimate the chance probability of
an interloper at less than 10\%, and suggest a more radical hypothesis.
They propose that the material in the Tytler clouds has
undergone significant stellar processing
which has reduced the deuterium abundance by a factor of ten.
Because the abundance of heavy elements in the Tytler clouds suggests little
stellar processing has occurred, Rugers and Hogan argue
the cloud is very inhomogeneous and the heavy elements have been dispersed.
This is not implausible since the material giving rise to
the absorption lines amounts to only tens of solar masses.

Should the primeval deuterium
abundance be high, the inferred baryon density is very low,
around 1\% for a Hubble constant of $70\kms\Mpc^{-1}$, making the
case for exotic, nonbaryonic dark matter ironclad as
all estimates for the total matter density exceed 10\%, and many
by a wide margin.  However, a high primeval D abundance requires
that more than 90\% of the
material in the local ISM has been though stars at least once
since measurements made by the Hubble Space Telescope$^7$ indicate that
(D/H)$_{\rm ISM}$ $=(1.6\pm 0.1)\times 10^{-5}$.
This runs counter to many models of the chemical evolution
of our galaxy and is strongly constrained by Gloeckler and Geiss'
$^3$He measurement.

Deuterium is first burnt to $^3$He, which, according
to conventional stellar models, is much more difficult to destroy,
suggesting that a high primeval deuterium abundance should
be reflected in a high $^3$He abundance in the ISM.
The measurement by Gloeckler and Geiss of the
$^3$He abundance in the local ISM, ($^3$He/H)$_{\rm ISM}
= (2.1^{+0.9}_{-0.8}) \times 10^{-5}$, precludes this possibility.
Further, the sum of the D and $^3$He ISM abundances is essentially
equal to that inferred for the solar system 4.5\,Gyr ago,
[(D+$^3$He)/H]$_\odot$ $=(4.2 \pm 1) \times 10^{-5}$, suggesting stars
increase $^3$He by burning D, but do not otherwise significantly
deplete or produce it.

Conventional stellar wisdom also predicts that the sum of
D + $^3$He should grow significantly with time due to $^3$He
production by low-mass stars$^8$.  The measurement by Gloeckler and Geiss is
inconsistent with this prediction.  While their measurement has
shed some light on the chemical evolution of $^3$He,
it has also cast doubt on the conventional wisdom.  At this moment,
any argument for or against a particular value of the primeval
deuterium abundance based upon the chemical evolution of $^3$He
holds little weight.

There are also pieces missing in the $^7$Li story.  Its abundance
is measured in the atmospheres of the oldest stars in the halo of
the Galaxy, ($^7$Li/H) $= 1.5 \pm 0.5 \times 10^{-10}$.  If this
is the primordial value, it favors a low baryon density (high
deuterium).  However, some stellar models indicate
that these stars could have reduced
their initial $^7$Li abundance by a factor as large as two,
in which case it could be consistent with a high baryon density
(low primeval deuterium).

The Keck Telescope makes it a good bet that within
a few years the primeval deuterium abundance will be determined unambiguously,
and with it the value of the baryon density.  An important
check may come a few years later from detailed mapping of the
anisotropy of the cosmic background radiation which provides an
independent way of determining the baryon density to similar precision$^9$.
Once the baryon density is known, nuclear physics in the early Universe
can used to teach us about the ``chemistry'' of $^3$He and
$^7$Li in the interstellar medium.

\bigskip
\noindent{\it David N. Schramm and Michael S. Turner are Professors in
the Departments of Astronomy \& Astrophysics and Physics at The University
of Chicago, Chicago, IL~~60637-1433, USA.}

\newpage

\end{document}